**Human-LLM Collaborative Inductive Coding for Conceptualizing K-12 Educator AI Use**

Alex Liu[a*], Min Sun[a], Lief Esbenshade[a], Michael Xiao[a], Victor Tian[a], Zachary Zhang[b], Kevin He[b]

**Affiliations**

[a] College of Education, University of Washington, Seattle, WA, United States

[b] Colleague AI

**Corresponding author**

Alex Liu, University of Washington, College of Education, Seattle, WA 98195, United States. Email: alexliux@uw.edu

**Abstract**

Qualitative researchers increasingly encounter interaction corpora whose scale exceeds what manual coding alone can address, and large language models (LLMs) are frequently proposed as analytic assistants. The open questions are not whether LLMs can participate in qualitative analysis but to what extent, in what phases, and under what safeguards. This article provides a detailed procedural account of a multi-phase human-LLM collaborative pipeline that adapted open, axial, and selective coding to develop a hierarchical codebook from 45,000 messages exchanged between K-12 educators and a generative AI platform. Across three phases, LLMs generated candidate labels and structured annotations at scale, while human researchers retained conceptual authority over category definitions, merging decisions, and interpretive frameworks. The resulting instrument was then tested through systematic human coding, in which three trained coders with educational domain expertise applied the codebook to an independent sample of 2,560 messages, established reliability through iterative calibration using set-valued agreement measures appropriate for multi-label annotation, and extended the instrument with five codes that the LLM-assisted phases had not surfaced. The final codebook comprises 72 items within 19 categories and six domains. We reflect on the methodological decisions the pipeline required, including the choice of a conversational unit of analysis, the treatment of the LLM as a labeling instrument rather than an interpretive agent, the measurement of intercoder agreement under multi-label coding, and the conditions under which human domain expertise remained decisive. The account is offered as an auditable template for qualitative researchers considering LLM assistance in codebook development while preserving human interpretive authority.

## Introduction

Qualitative methodology has long balanced two commitments that sit in tension, depth of interpretation and breadth of evidence. Interaction log data has intensified this tension. Digital platforms now record authentic professional and personal activity at a volume that invites qualitative questions while overwhelming qualitative capacity. The corpus examined in this study is a case in point. An AI-powered platform built for education accumulated conversations from more than 15,000 active K-12 educators, producing a record of authentic professional practice that no research team could read in full, let alone code line by line. Existing evidence on how educators use generative AI rests largely on self-reported surveys or small interview samples (Cheah et al., 2025; Esbenshade et al., 2025; Zaimoğlu & Dağtaş, 2025), in part because the interaction data that would support stronger claims resist conventional analysis at scale.

Large language models have been proposed as one response to this impasse. A growing body of work reports that LLMs can support theme discovery, deductive code application, and thematic analysis across qualitative corpora (Chew et al., 2023; De Paoli, 2024; Gao et al., 2025; Shin, 2025; Tai et al., 2024; Wang et al., 2025). The proposal remains contested. Some researchers warn that delegating analysis to a model threatens the interpretive foundations of qualitative inquiry (Friedman et al., 2024) or can introduce systematic bias into the resulting claims (Ashwin et al., 2025), and interview evidence suggests that qualitative researchers themselves hold divided and often ambivalent views about where model assistance belongs (Schroeder et al., 2025). The literature also remains dominated by demonstrations on small or benchmark datasets, concentrated on whether model output matches human output on a fixed coding task. Far less attention has been given to the fuller methodological problem that practicing qualitative researchers actually face, which is how to move from an unlabeled corpus to a validated, hierarchical coding instrument while preserving the interpretive authority that gives qualitative work its warrant. Codebook development is precisely the phase of qualitative research in which conceptualization, judgment, and familiarity with the data matter most, and it is therefore the phase in which claims about LLM assistance require the most careful specification.

We take the position that the productive question is not whether LLMs can substitute for human coders but to what extent, and in what areas, LLM assistance can be incorporated into established inductive procedures without weakening them. Answering that question requires a documented case in which the division of labor between human and model is explicit at every phase, in which the resulting instrument is subjected to independent human application, and in which the points of human intervention are reported rather than smoothed over. This article provides such a case. Two methodological questions guided the work.

**RQ1.** How can established inductive coding procedures, specifically open, axial, and selective coding, be adapted into a human-LLM collaborative pipeline for developing a hierarchical codebook from a large corpus of educator-AI interaction data?

**RQ2.** To what extent, and in what areas, does the resulting instrument hold up when applied independently by trained human coders, and what does systematic human coding contribute that the LLM-assisted phases do not?

The empirical setting is a corpus of 45,000 educator and AI messages drawn from the platform described above. Through the three-phase pipeline reported here, we developed a hierarchical codebook capturing the breadth of educator AI use across six professional domains, then validated and extended the instrument through human coding of an independent sample of 2,560 messages. The substantive patterns of educator AI use documented with this instrument are briefly reported. The present article concentrates on the method, because we found that the methodological decisions the pipeline required, concerning units of analysis, prompt design, label consolidation, agreement measurement, and the boundaries of model assistance, are the decisions other research teams will face and are rarely reported in enough detail to be reused.

The remainder of the article proceeds as follows. We first situate the work within the literatures on codebook development and LLM-assisted qualitative analysis. We then describe the data source and the three LLM-assisted coding phases, followed by the human coding phase through which the instrument was validated and extended. We close by discussing the methodological lessons of the pipeline, the conditions under which human expertise remained decisive, and the limitations of the approach.

## Background

### Inductive Coding and Codebook Development

The pipeline reported here is grounded in the family of inductive procedures associated with grounded theory and its descendants. Open coding surfaces initial concepts through close reading and labeling of data segments, axial coding relates and consolidates those concepts into categories through attention to conditions, context, and consequences, and selective coding integrates categories into a coherent structure (Charmaz, 2006; Corbin & Strauss, 1990, 2014). When the analytic goal is a reusable instrument rather than a theory, these procedures culminate in a codebook, a structured set of code definitions with inclusion and exclusion criteria that supports consistent application by multiple coders (Saldaña, 2021). Codebook quality is conventionally evidenced through intercoder agreement, yet standard chance-corrected coefficients presuppose single-label judgments and penalize partial overlap between coders' code sets as though it were full disagreement, a poor fit for multi-label coding of compound professional activity (Passonneau, 2006). These commitments, phased inductive development, explicit code definitions, and agreement evidence suited to the coding task, define the methodological standard against which any LLM-assisted variant should be judged.

**LLM Assistance in Qualitative Analysis**

The question of whether computational methods can share in qualitative coding predates generative models. Nelson et al. (2021) compared hand coding with three families of computer-assisted text analysis and concluded that algorithmic methods could augment but not replace the interpretive work of human coders. Than et al. (2025) revisited that comparison for generative LLMs and reached a parallel judgment, finding that generative models change the economics of coding while leaving the locus of interpretation with the researcher. Within this lineage, recent studies have explored several points of entry into the analytic workflow. De Paoli (2024) performed an inductive thematic analysis of semi-structured interviews with a language model and characterized both the plausibility of the generated themes and the limits of the approach, including sensitivity to segmentation choices imposed by token constraints. Chew et al. (2023) developed LLM-assisted content analysis for deductive coding, and Dunivin (2024) reported that chain-of-thought prompting can match human performance on some hermeneutic coding tasks while falling short on others. Wang et al. (2025) piloted LLM-assisted thematic analysis of online social network data, Shin (2025) examined ChatGPT-assisted co-coding of classroom dialogue from the perspective of a single researcher, and Gao et al. (2025) built tooling that structures the model's analytic steps into a controllable reasoning chain. In education research specifically, Barany et al. (2024) explored LLM support for qualitative codebook development, and Liu and Sun (2025) examined the validity of LLM-assisted textual analysis of policy stakeholder interviews against human coding.

Methodological work has also begun to formalize how LLM assistance should be structured and judged. Tai et al. (2024) proposed treating repeated LLM coding runs as independent coders and positioned the model as a complement to, rather than a replacement for, human reliability procedures. Misra et al. (2026) compared traditional and researcher-interpreted LLM approaches to the same corpus, and Xu (2026) examined the practices, ethics, and reflexivity of thematic analysis conducted with generative AI. Alongside these constructive accounts sits a body of sustained caution. Friedman et al. (2024) argued that AI assistance risks hollowing out the researcher's engagement with data, Ashwin et al. (2025) demonstrated that LLM-based analysis can introduce systematic bias into downstream claims, Kristensen-McLachlan et al. (2025) found that model annotation reliability is uneven across tasks, and Matta et al. (2026) proposed procedural safeguards for protecting qualitative rigor. Read together, these literatures converge on two observations relevant here. Models are most dependable when the task is bounded, such as proposing candidate labels or applying a fixed scheme, and least dependable when the task requires sustained interpretive judgment across a corpus. In addition, design decisions that precede any model call, including how the data are segmented and what the prompt asks for, substantially shape what the model can contribute.

Three gaps in this literature motivated the present account. Most published demonstrations operate on corpora small enough that full human analysis remains feasible, which leaves untested the setting where LLM assistance is most consequential. Most address a single analytic phase, whereas codebook development runs from unstructured exploration through structured consolidation, and the appropriate division of labor plausibly differs by phase. Finally, few studies subject the LLM-assisted product to independent application by trained human coders on new data, which is the test that matters if the product is meant to function as an instrument. The pipeline described below was designed with these gaps in view.

# Method

To investigate how K-12 educators use generative AI for instructional and professional purposes, we developed a multi-phase inductive coding pipeline that integrates established qualitative coding techniques with selective use of LLMs. The primary methodological output is a hierarchical codebook that captures the breadth of educator AI use. In keeping with the reflexive commitments of qualitative research, we report not only the procedures but the points at which human judgment intervened and the rationale for each design decision.

## Data Source

This study draws on educator-AI interactions collected from [redaction], an open-registration generative AI platform designed to assist educators with instructional planning, content development, assessment, differentiation, and professional communication. At the time of data collection, the platform supported over 15,000 active users, including K-12 teachers, school administrators, paraprofessionals, and instructional specialists. All usage was voluntary and initiated by educators as part of their authentic professional work.

The platform allows educators to submit free-form natural language prompts and receive AI-generated responses, including instructional materials, pedagogical strategies, and communication artifacts. Prompts span a wide range of grade levels (K-12 and educator preparation programs), subject areas (literacy, STEM, social studies, arts, career and technical education), and instructional contexts (multilingual learners, varied pacing needs, diverse classroom configurations). Prompt types range from brief single-turn requests to complex multi-turn exchanges involving iterative revision, format adaptation, and pedagogical elaboration. This dialogic structure preserves the evolving nature of instructional decision-making, as educators refine or redirect AI outputs to better align with student needs, lesson goals, or contextual constraints. Each message was accompanied by metadata (conversation ID, timestamp, platform feature of origin, and anonymized user ID), enabling both message-level and conversation-level analysis.

The corpus analyzed across the phases reported below totals 45,000 educator and AI messages. This figure comprises the 21,147 conversation trios used across the open, axial, and selective coding phases, together with the 2,560 messages in the human-coded validation sample. Trios were drawn as overlapping three-message windows within each conversation, where the follow-up message that closes one trio serves as the opening request of the next, so consecutive trios within a conversation share messages rather than each contributing three unique messages to the total. The inductive samples were drawn from the full platform corpus available at the time of each coding phase. The validation sample was drawn as a stratified random sample by month across the June 2025 to June 2026 window, and the two samples are independent at the message and conversation level, since no message or conversation appearing in the validation sample also appears among the inductive coding samples. This independence supports using the validation sample to test codebook coverage and reliability, and it additionally allows an assessment of coding stability over the full year the validation sample spans.

**Ethical considerations.** All data were collected under the platform's terms of use, which explicitly stated that de-identified data could be used for research purposes. Personally identifiable information was removed prior to analysis. The study protocol was reviewed and approved under educational research guidelines by the Institutional Review Board at [Institution redacted for peer review]. All analysis using commercial LLMs occurred through enterprise API accounts with contractual terms confirming that submitted data would not be used for model training. The human subjects of this study are the platform's educator users. The data analyzed consist of educators' messages and the AI's responses to them. Some educator messages include excerpts of student writing that a teacher chose to paste into their own request, for example when asking for feedback on a piece of student work. This content is analyzed as part of the educator's submitted message rather than as separately collected student data, consistent with the IRB determination that the study concerns educator platform use rather than student participants.

## Overview of the Inductive Coding Pipeline

Codebook development followed a three-phase inductive process grounded in established qualitative methodology (Charmaz, 2006; Corbin & Strauss, 2014). Open coding served initial theme discovery, axial coding served category refinement and relationship mapping, and selective coding served final consolidation. At each phase, human researchers retained conceptual authority over category definitions, merging decisions, and interpretive frameworks, while LLMs were employed to support theme generation at scale and structured label extraction across large numbers of messages. Figure 1 summarizes the full pipeline, from data sampling through the three coding phases to human validation.

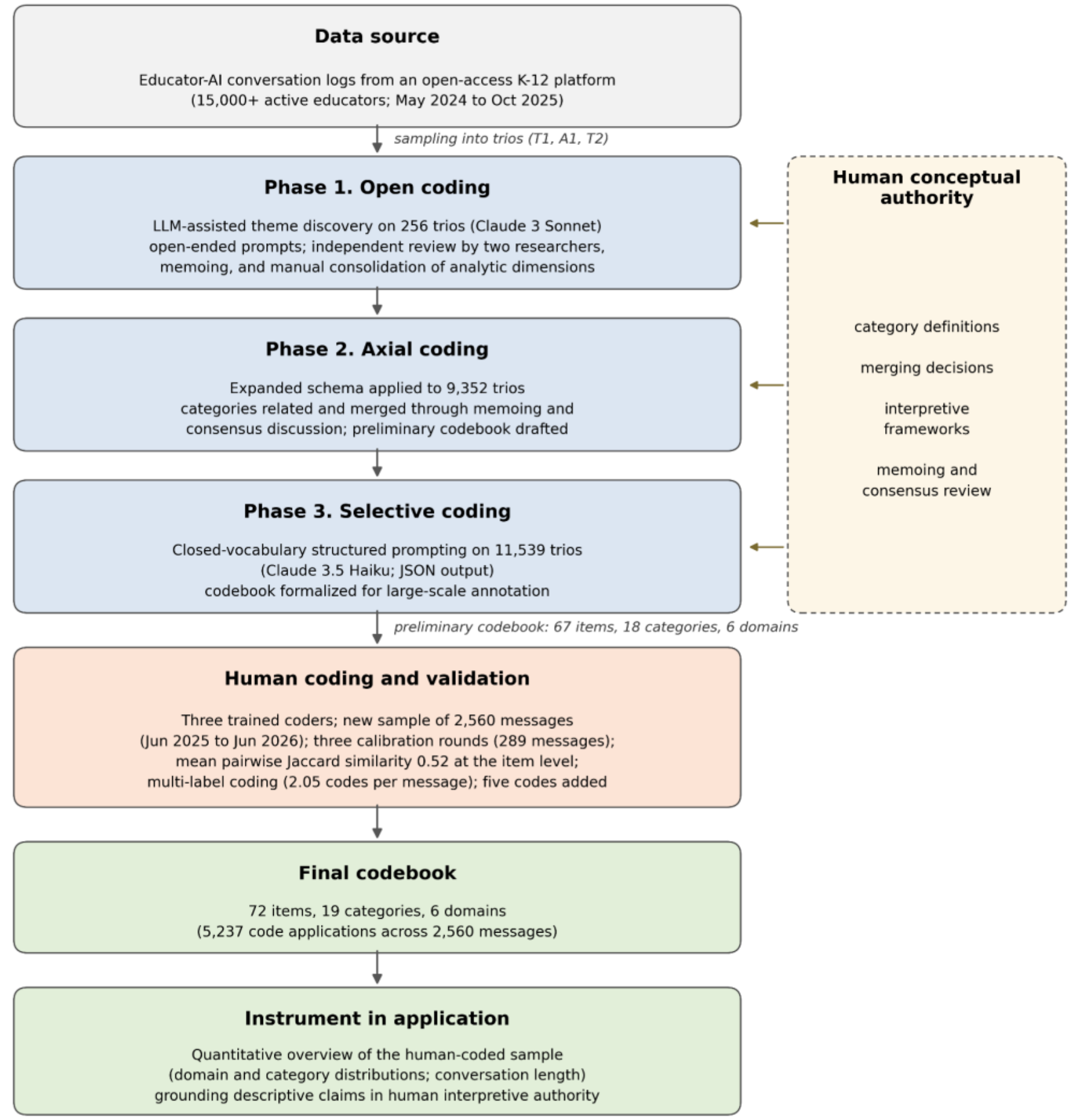

**Figure 1.** Workflow of the human-LLM collaborative inductive coding pipeline. Blue boxes denote LLM-assisted phases, the orange box denotes human coding and validation, and green boxes denote outputs. The dashed lane indicates that human researchers retained conceptual authority over category definitions, merging decisions, and interpretive frameworks throughout the LLM-assisted phases.

## Open Coding for Inductive Theme Discovery

The first stage of analysis involved open coding to inductively explore the education-related themes embedded in educator-AI interactions. Drawing on established qualitative coding practices (Charmaz, 2006; Corbin & Strauss, 1990), this phase aimed to surface recurring patterns and instructional constructs that would inform the development of a hierarchical codebook.

A central methodological consideration in qualitative analysis is defining the unit of analysis, the segment of data that is interpreted, labeled, and compared. There is no single best approach. Wang et al. (2025) and Shin (2025) provided entire documents to LLMs to preserve context, Gao et al. (2025) pre-clustered raw qualitative text based on semantic similarity, and De Paoli (2024) segmented documents into smaller chunks due to token limitations. In our case, single educator-AI conversations could reach considerable length, sometimes exceeding 500 messages. To balance contextual richness with token constraints, we began with triadic units, each consisting of an educator prompt (T1), the AI response (A1), and the educator's follow-up (T2), if present. These trios were treated as self-contained instructional interactions, enabling us to surface dialogic patterns while preserving coherence and instructional intent.

We employed claude-3-sonnet-20240229 (Claude 3 Sonnet) for this phase. To assess the validity of using the model for theme discovery in this context, we began with a small test sample of 256 randomly selected trios. Each trio was submitted individually to the Anthropic API, accompanied by a structured prompt designed to elicit information characterizing the diverse ways educators use the AI tool in their daily work. Because no session memory persisted across API calls, each trio was interpreted independently, eliminating carryover effects. The prompt, reproduced in full in Appendix A, asked the model to identify the content subject domain, pedagogical strategies, the degree to which the request addressed instructional strategy, and the specific support the educator requested, returning results in a strict JSON structure. Table 1 presents examples drawn from the raw LLM outputs.

**Table 1.**

*Examples of educator requests and corresponding LLM open coding results.*

| Request | Content domain | Pedagogical strategies | Score | AI tool support |
|---|---|---|---|---|
| I need to think of an introductory unit idea for a college and career prep class for a 9th grade class | College and career preparation | None | 2 | Idea generation |
| The unit needs to be based on skills needed to be successful for high school | Study skills | None | 3 | Lesson design |
| How do I prank my co-workers? | None | None | 1 | Idea generation |
| Can you make a rubric for the mind-mapping activity above? | Social studies / cultural studies | Rubric creation | 3 | Rubric creation |
| Thank you for providing the learning target K.NS.3. This will help me craft an authentic assessment aligned with that standard. | Math / counting and cardinality | Formative assessment | 3 | Assessment creation |

After the LLM generated analytic summaries for all trios, two researchers independently reviewed the outputs to assess their accuracy and interpretive validity. Discrepancies were resolved through discussion and collaborative memo writing, followed by the consolidation of analytic dimensions and an initial categorization of labels. Prior studies (e.g., Shin, 2025) have incorporated interfaces such as ChatGPT to assist in categorizing LLM-generated labels. At this stage, with the relatively small size of our trial sample, we completed this step manually to deepen our human understanding of the data.

Based on this validation and synthesis process, we refined our labeling categories to include extended contextual information, such as references to student needs and pedagogical frameworks mentioned in educator prompts, and identified recurring instructional tasks as examples. These refined categories were then introduced to the LLM to support inductive coding at scale in subsequent phases of analysis.

**Axial Coding for Conceptualization and Codebook Development**

To develop a fuller understanding of educator-AI conversations, the second stage scaled up the analysis to support comprehensive codebook construction. We expanded the dataset to 9,352 trios, retaining the trio format to preserve dialogic and instructional context while deepening theme discovery. This phase drew on the logic of axial coding as described by Corbin and Strauss (2014), wherein researchers relate codes by identifying underlying conditions, contextual factors, actions, and consequences. Our goal was not to build theory but to translate emergent patterns into a reproducible framework suitable for large-scale annotation of educational dialogue. The final output of this stage was a preliminary codebook that captures the instructional and professional complexity embedded in educator-AI interactions. Throughout this process, researchers continued memoing to document rationales for code merges, disagreements, and iterative refinements.

In this phase, the LLM was instructed to freely generate responses for each analytic dimension. This approach reduced variability in label naming across similar concepts, while preserving the opportunity for surfacing new or unexpected categories. The complete prompt is included in Appendix B. For the request "How can I create culturally relevant activities for my Grade 3 fractions lesson plan, where a majority of my students are Spanish speakers," for example, the model returned the content area Math/Fractions, the grade level Grade 3, the contextual information that a majority of students are Spanish speakers, the instructional task of incorporating real-world examples, and the requested support of creating culturally relevant activities.

As in the open coding phase, two researchers randomly selected a sample of 1,000 LLM-generated outputs to review for accuracy and interpretive quality. Discrepancies in conceptualizing labels were resolved through collaborative meetings and memo-based synthesis. After validation, outputs for each analytic dimension were returned to the LLM to propose label unifications and identify common representative terms. Researchers then clustered LLM-generated labels within each analytic dimension by grouping them based on semantic similarity, pedagogical intent, and frequency of use. For example, terms such as “differentiated strategies” and “differentiated support” were merged, as were “create quiz” and “generate formative assessment.” Adjacent but distinct constructs such as “inquiry-based learning” and “project-based learning” were intentionally preserved. This human-in-the-loop process ensured that the emerging codebook maintained conceptual clarity while staying grounded in authentic instructional language.

The outcome of axial coding was a preliminary codebook with pre-determined labels for categories including Educational Context, Content Focus, and Instructional Practices (Appendix C). The Pedagogical Frameworks dimension remained open-coded to allow for emergent variation. To assess the coverage and efficiency of this initial codebook and to explore the relationships among categories and labels, it was incorporated into the prompt design for the next stage of analysis.

**Selective Coding for Structured Prompting and Codebook Refinement**

With the preliminary codebook established, the selective coding phase focused on validating, refining, and structurally formalizing the codebook for broader application. This stage marked the transition from open-ended annotation to a closed-vocabulary approach, wherein predefined labels were applied to a new set of 11,539 educator-AI message trios. Unlike earlier phases that emphasized exploratory theme generation, the LLM was now prompted to select from existing codebook categories and return outputs in structured JSON format, enhancing consistency and enabling efficient downstream parsing (see Appendix C for the complete prompt). We employed claude-3-5-haiku-20241022 (Claude 3.5 Haiku) for the remainder of the analysis, as this model became available during the study period.

Because the coding LLM functions as a labeling instrument rather than an interpretive one, the change of model between phases does not threaten the validity of the analysis in the way it would if model output were treated as a finding in itself. At every phase, the LLM's role was confined to generating initial candidate labels and structured annotations. Human researchers then inspected, discussed, consolidated, and reinterpreted these outputs before any category, merge, or classification entered the codebook. The model transition therefore changes which system proposed a given label, not who determined whether that label was retained, refined, or discarded. We treat this human-in-the-loop structure, rather than cross-model output agreement, as the relevant safeguard against instrument-induced artifacts, and we return to this point in the Discussion when addressing reflexivity about LLM-assisted coding more broadly.

This phase served three primary goals. The first was to evaluate the completeness and usability of the preliminary codebook at scale. The second was to assess the model's reliability in applying structured codes under constraint. The third was to identify underrepresented, misclassified, or emergent instructional patterns requiring refinement. To support the first goal, we incorporated validation through "Other" annotations and human review. To maintain inductive flexibility, the model was permitted to select "Other" when no predefined option was appropriate, with an accompanying textual justification. Human researchers manually reviewed all "Other" responses to determine whether recurring patterns merited new codes. For example, emergent instructional contexts such as homeschooling and afterschool programs, which initially emerged as labels under "Other," were later incorporated into the codebook under the Student Needs and Context domain.

These steps mirror the selective coding process in traditional qualitative analysis and follow the procedures described by Saldaña (2021). The final output of this phase was a structured codebook, developed during the coding process to represent meaningful categories and their interrelationships through code pruning and consolidation. To enhance analytic tractability, the research team evaluated code frequency and semantic distinctiveness across the structured annotations. Low-frequency codes that lacked conceptual clarity or were frequently misapplied were either merged with related categories or removed. For example, "Gallery Walk" was folded into broader subcategories within Collaborative Learning, and the underutilized "Learning Progression" category was eliminated due to redundancy and inconsistent usage. These pruning decisions were guided by code frequency thresholds, coder feedback, and alignment with existing pedagogical literature.

Through iterative reading, comparison, and contrast of coded outputs, the team arrived at a hierarchical codebook structure (**Domain**, *Category*, "Item") that organized broad professional responsibilities and discourse markers, while also capturing fine-grained pedagogical practices and instructional tools. The resulting codebook achieved a balance of domain relevance, internal coherence, and scalability for large-scale analysis. At the close of this phase it comprised six top-level domains, 18 mid-level categories, and 67 fine-grained instructional items. Table 2 provides illustrative categories and examples of items under each domain, and the full instrument is available as described in the Data Availability statement.

**Table 2.**

*Domains in coded educator-AI conversations and example categories and items.*

| Domain | Category | Example item |
|---|---|---|
| Instructional practices | Critical thinking and inquiry | "Historical thinking" |
| | Explicit teaching | "Modeling problem solving" |
| | Project-based and real-world learning | "Real-world engagement and scenarios" |
| | Collaborative learning | "Group work" |
| | Instructional routine | "Learning progression and routine adjustments" |
| | Engagement and motivation | "Actionable engagement strategy" |
| | Differentiation and accessibility | "Tiered scaffolding"; "Multilingual learner support" |
| Curriculum and content focus | Planning | "In-class activity design and adjustment" |
| | Technology integration | "Multimedia use for instruction" |
| Student needs and context | Classroom setting | "Low-tech environments" |
| | Student profile | "Special education (IEP/504)" |
| | Career readiness | "Student career exploration" |
| Assessment and feedback | Assessment | "Generate formative assessment" |
| | Feedback | "Data-driven student learning progress monitoring" |
| Professional responsibilities | Professional development | "Reflection on teaching practices" |
| | Communication | "Administrative messaging" |
| Other | Discourse continuity | "Format modification" |
| | Non-educational queries | "Non-educational queries" |

In addition to structured categorical codes, open-text metadata fields were retained for subject area, grade level, and explicitly mentioned pedagogical frameworks, allowing for future stratified or filtered analyses. This stage completed the LLM-assisted portion of codebook development. The instrument's standing as a qualitative coding tool, however, rests on the human coding phase described next.

## Codebook Validation Through Human Coding

We anchored codebook development in validation through human coding. Although the previous steps indicated that the LLM-assisted inductive coding approach produced a hierarchical codebook structure with 67 items across six domains, the instrument underwent further refinement when it was systematically applied by human qualitative coders with educational domain knowledge. This refinement was conducted by three trained researchers, who applied the codebook to a newly collected random sample of 2,560 educator messages from June 2025 to June 2026. This process both validated the existing codes and revealed gaps in the instrument's coverage as educator AI use evolved from the early adoption stage captured in the inductive coding sample to new conversations collected over a period of one year. Most importantly, final judgment about codebook validation and application remained grounded in human domain expertise.

### Coders, Calibration, and Agreement

All three coders had substantial educational domain knowledge. One held a doctoral degree in education, one held a master's degree in education, and one had an undergraduate minor in education along with industry research experience. Prior to independent coding, the team completed three iterative calibration rounds using 289 messages as a hold-out set, coded by all three coders, to establish intercoder reliability and align the coders' interpretive frameworks. Reliability improved across the three calibration rounds through collaborative discussion and the development of a shared coding protocol.

The coding was multi-label over a large codebook instead of the single-label classification setting that conventional agreement statistics assume. Standard chance-corrected coefficients such as Krippendorff's alpha are built for single-category judgments and penalize partial overlap between coders' code sets as though it were full disagreement, even when the coders' interpretations substantially converge (Passonneau, 2006). We therefore report agreement using Jaccard similarity between coders' code sets, a measure suited to set-valued annotation, computed pairwise across the three coders on the 289-message overlap set. Mean pairwise Jaccard similarity was 0.52 at the codebook item level, ranging from 0.46 to 0.55 depending on the granularity at which agreement is assessed (item versus category versus domain), indicating that coder pairs typically shared at least one assigned code and frequently the full code set. Calibration discussions indicated that disagreement arose primarily at boundaries between semantically adjacent categories rather than in core structural distinctions between domains, consistent with the pattern of higher agreement at coarser code granularity. Remaining ambiguities were resolved through structured disagreement discussions, which also informed the codebook refinement described below.

Of the full 2,560-message validation sample, 289 messages (11.3%) were independently coded by all three coders as the calibration and reliability check. The remaining 2,271 messages (88.7%) were distributed across coders for single-coder annotation following calibration. Messages were multi-labeled, with educator prompts receiving an average of 2.05 codes per message (5,237 code applications across 2,560 messages). This multi-label approach reflects the compound nature of instructional work. A single prompt frequently integrates planning with differentiation, or assessment design with standards alignment.

**Codes Added During Human Coding**

During calibration, coders encountered messages that fell clearly within the scope of the codebook's domains but could not be adequately captured by any existing item. These recurring gaps, identified through structured disagreement discussions and codebook boundary analysis, motivated the addition of new codes. Five codes were added during the human coding phase, bringing the total from 67 to 72 items and from 18 to 19 categories.

**"Culturally Responsive Teaching"** (**Instructional Practices**, *Differentiation and Accessibility*). Messages in which educators designed instruction that drew on students' cultural backgrounds, funds of knowledge, or community contexts appeared with sufficient frequency to warrant a distinct code, rather than being subsumed under the broader Differentiated Instructional Strategies item.

**"Foreign Language Skill Development"** (**Instructional Practices**, *Explicit Teaching*). A cluster of messages from world language teachers focused specifically on building target-language proficiency. These interactions had a distinct pedagogical character (vocabulary acquisition, grammar practice, conversational fluency) that the existing ELA Skills Development code did not adequately capture.

**"Classroom Management"** (**Student Needs and Context**, *Classroom Setting*). While the original codebook included Student Behavioral Intervention for reactive strategies, coders identified a set of proactive classroom management interactions (routines, procedures, transitions, expectations) that addressed the organizational structure of learning environments rather than individual student behaviors.

**"Class, School, or Community Event"** (**Professional Responsibilities**, *Communication*). Educators frequently sought help planning or communicating about events (assemblies, family nights, field trips, college fairs) that did not fit within existing Professional Communication or Administrative Documentation codes.

**"School District Finance" (Professional Responsibilities**, *Management*). A small but distinct set of messages from school administrators addressed budgeting, resource allocation, and financial planning, representing a professional context not captured by existing codes focused on instructional or communication tasks.

These additions are methodologically informative beyond their content. Each captures either a population of platform users, a professional context, or an evolution in use that the LLM-assisted phases had not surfaced, despite those phases processing roughly eight times as many trios as the human coders read messages. Sustained human application of an instrument to new data detected coverage gaps that scale alone did not.

### The Final Codebook

The final instrument comprises 72 items organized within 19 categories and six domains (Full codebook is included in online repository with link available in Data availability section). Each item includes a description with explicit inclusion and exclusion criteria. This codebook represents the consolidated output of both the LLM-human collaborative inductive development process, which established the conceptual structure and initial item set, and the domain-expertise-grounded deductive application process, which tested, refined, and extended the instrument through systematic coding.

## The Instrument in Application

To characterize what the validated instrument captures, we summarize code frequencies from the human-coded validation sample of 2,560 messages, which yielded 5,237 code applications under the multi-label protocol described above. We report these distributions at the domain and category levels and present them as a description of the validation sample rather than as statistically generalizable estimates for the full corpus. The two samples serve distinct purposes within the qualitative design. The large inductive sample supports conceptualization, meaning the development of the codebook and the analytical vocabulary, while the human-coded sample validates the instrument and grounds descriptive claims in human interpretive authority.

At the domain level (Table 3), Instructional Practices accounted for the largest share of code applications (1,913; 36.5%), followed by Curriculum and Content Focus (909; 17.4%), Assessment and Feedback (603; 11.5%), Professional Responsibilities (450; 8.6%), and Student Needs and Context (396; 7.6%). The Other domain, which combines Discourse Continuity and Non-Educational codes, accounted for 924 applications (17.6%), reflecting the conversational maintenance work of format changes and continuation requests, together with occasional off-task use, that accompanies substantive instructional interaction.

**Table 3.**

*Domain-level distribution of code applications in the human-coded validation sample (2,560 messages; 5,237 code applications).*

| Domain | Code applications | % of total |
|---|---|---|
| Instructional Practices | 1,913 | 36.5 |
| Curriculum and Content Focus | 909 | 17.4 |
| Other | 924 | 17.6 |
| Assessment and Feedback | 603 | 11.5 |
| Professional Responsibilities | 450 | 8.6 |
| Student Needs and Context | 396 | 7.6 |

At the category level (Figure 2), Planning was the single most frequently coded category (777; 14.8%), followed by Explicit Teaching (646; 12.3%), Discourse Continuity (561; 10.7%), Assessment (465; 8.9%), and Differentiation and Accessibility (396; 7.6%). The remaining categories each accounted for less than 5% of applications. The prominence of Discourse Continuity reflects the multi-turn character of the interactions, as educators frequently issued requests to reformat, extend, or adjust material generated earlier in a conversation. The multi-label protocol means that percentages describe shares of code applications rather than mutually exclusive message counts, since a single message frequently combined codes from more than one category.

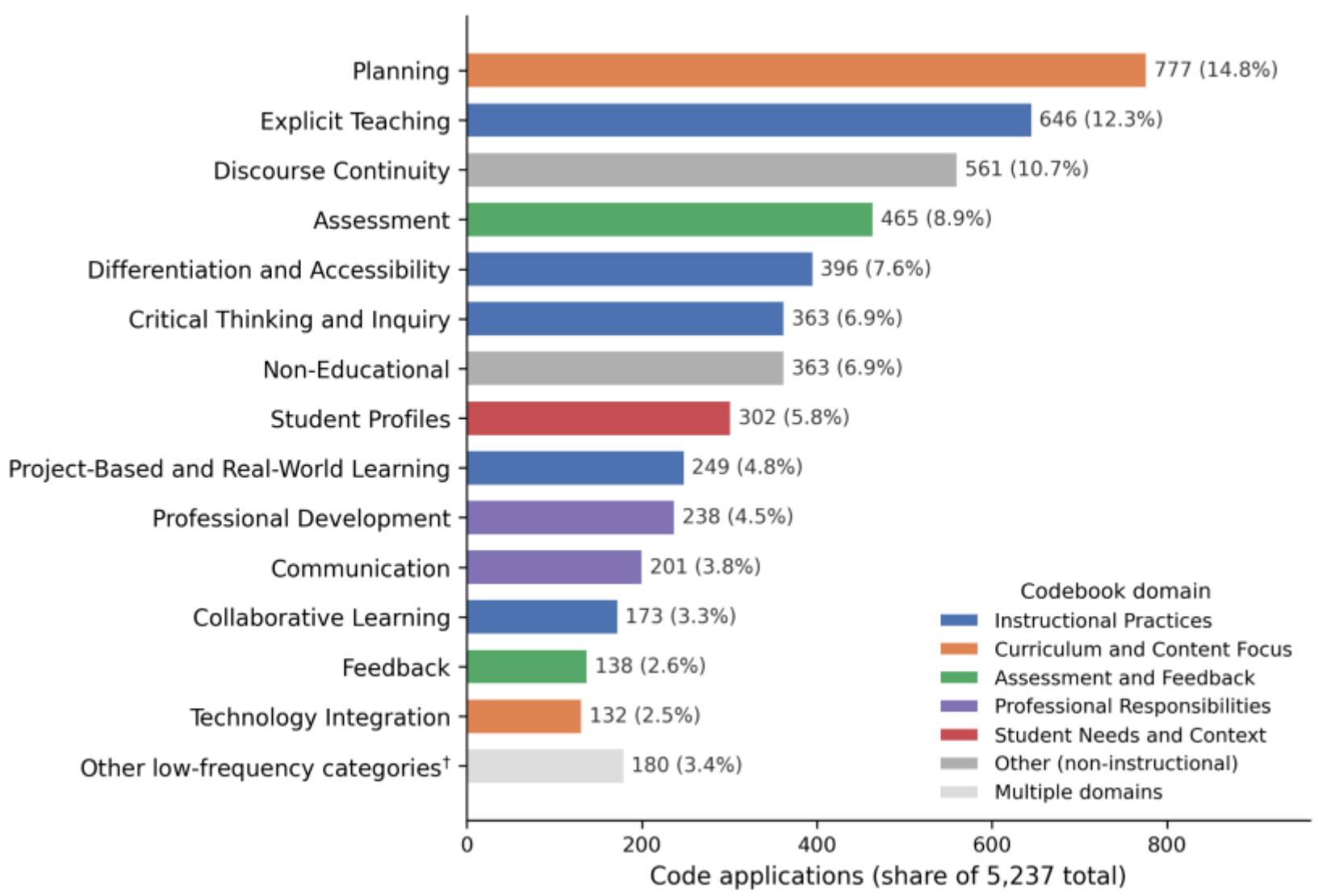

**Figure 2.** Category-level distribution of code applications in the human-coded validation sample (2,560 messages; 5,237 code applications), colored by codebook domain. Discourse Continuity and Non-Educational are non-instructional categories and are shaded gray. Four categories that each account for less than 2% of applications (Classroom Setting, Engagement and Motivation, Instructional Routine, and Career Readiness) are combined.

The validation sample's 2,560 messages are drawn from 500 distinct conversations, and conversation length varies considerably. In the sample, 291 conversations (58.2%) contain one to three messages, 124 (24.8%) contain four to six, 36 (7.2%) contain seven to ten, and 49 (9.8%) contain eleven or more, with a maximum of 68 (Figure 3). The dialogic structure that motivated the trio unit of analysis is thus a property of the data rather than an artifact of the design, and it underscores why units that sever educator turns from their conversational context would misrepresent the phenomenon being coded.

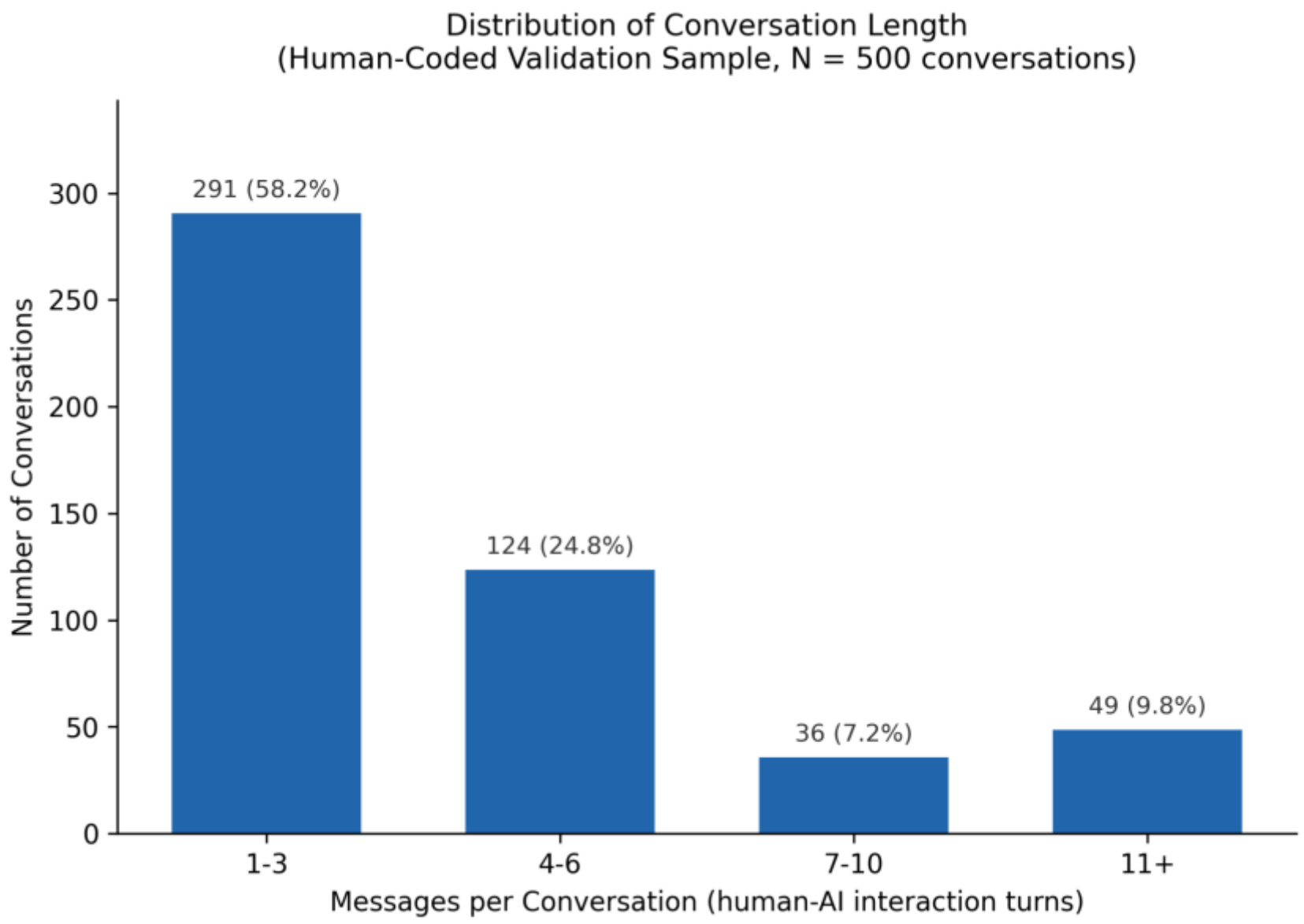

**Figure 3.** Distribution of conversation length, measured as the number of messages exchanged between educator and AI, in the human-coded validation sample (500 conversations).

## Discussion

The pipeline reported here yielded a validated 72-item instrument from a corpus that manual analysis could not have addressed in full. Its methodological interest, however, lies less in the product than in what the process reveals about the division of interpretive labor between researchers and language models. We organize the discussion around the decisions that other research teams will face, and we characterize the extent, and the areas, in which LLM assistance proved dependable in our case.

### The Division of Interpretive Labor

Across all three LLM-assisted phases, the model's role was confined to proposing, and the researchers' role was to dispose. The LLM generated candidate labels, structured annotations, and suggested unifications. Human researchers decided what entered the codebook, what merged with what, and what was discarded. This division held constant even as the model's task shifted from open generation toward closed-vocabulary application, and it is the feature of the design we regard as least negotiable. It is also what distinguishes the pipeline from delegating analysis to a model and auditing a sample of its output. Every category definition in the final instrument passed through researcher discussion, memoing, and consensus, so the interpretive chain of custody that qualitative methodology requires remains intact and reportable.

Framed against the concerns raised in the emerging literature, the pipeline treats the model in the way qualitative researchers have long treated other instruments and intermediaries, as a component whose contributions must be inspected rather than trusted (De Paoli, 2024; Tai et al., 2024). The question we posed at the outset, to what extent and in what areas LLM assistance can be incorporated without weakening inductive procedural rigor, receives a differentiated answer in our data. The model was dependable for breadth, for consistency of structured annotation, and for surfacing recurring surface patterns across tens of thousands of trios, tasks in which its lack of fatigue and uniform attention are genuine advantages. It was not a source of conceptual authority at any phase, and the validation results indicate that it could not have been. The five codes added during human coding, and the boundary disagreements that calibration surfaced, mark exactly the areas in which sustained human engagement with data does work that scaled annotation does not.

The design also carries a reflexive obligation that manual analysis does not. A model is not a neutral reader. It brings dispositions from its training that shape which labels it proposes and which phrasings it favors, and researchers who consolidate model-proposed labels are interpreting an interpretation (Cambo & Gergle, 2022; Xu, 2026). We addressed this in three ways. Researchers read raw data alongside model outputs at every phase rather than encountering the corpus only through the model's summaries. Memoing recorded not only category decisions but occasions when model proposals were judged misleading, generic, or skewed toward surface features. Finally, the human coding phase confronted the instrument with data no model had preprocessed, which is where several of its silences became visible. These practices do not eliminate the concern that model involvement can bias analysis (Ashwin et al., 2025), but they make the sites of potential bias inspectable, which is what reflexivity can realistically promise.

**The Unit of Analysis as a Methodological Commitment**

The choice of overlapping conversational trios was among the most consequential design decisions in the pipeline, and it was made on qualitative grounds in addition to the computational ones. Educator-AI interaction is dialogic, and an educator turn read without the response it answers loses its instructional meaning. The trio preserved enough context to code intent while remaining small enough for consistent model processing, and the overlap between consecutive trios kept conversational continuity visible. The conversation length distribution in the validation sample, where more than 40% of conversations extend beyond three messages, confirms that context-severing units would have misrepresented a substantial share of the corpus. We offer the trio not as a universal answer but as an illustration that unit-of-analysis decisions in LLM-assisted work remain interpretive decisions, to be argued from the structure of the phenomenon rather than defaulted from token limits.

**Measuring Agreement Under Multi-Label Coding**

Validation raised a measurement problem that LLM-assisted studies inherit from qualitative methodology but rarely confront. Multi-label coding with a large codebook does not fit the assumptions of the chance-corrected coefficients conventionally used to evidence reliability, and forcing it into that mold penalizes partial convergence as full disagreement (Passonneau, 2006). Reporting Jaccard similarity over code sets, at multiple levels of the code hierarchy, gave a truthful picture of where coders converged and where they diverged, and the granularity gradient in our results, with agreement rising from item to category to domain level, localized disagreement at semantic boundaries between adjacent codes rather than in the instrument's structure. We suggest that researchers validating large hierarchical codebooks report agreement at each level of the hierarchy, since a single coefficient at the finest granularity understates the reliability of the structure while a single coarse coefficient overstates the reliability of the items.

**Instrument Stability Across Model Change**

Commercial LLMs change on timescales much shorter than qualitative projects. Our pipeline crossed one such change, moving between model versions between the axial and selective phases. The design absorbed this transition because the model was positioned as a labeling instrument whose outputs were always subject to human disposition, so a change of model changed which system proposed labels, not who decided their fate. Research designs in which model output is itself the finding do not have this protection, and we regard the instrument stance, together with full archiving of prompts, as the practical safeguard available to teams working with models they do not control.

**What Human Coding Contributed**

The validation phase repaid its considerable cost in three currencies. It produced reliability evidence that no amount of LLM-internal consistency can supply, since the claim at issue is whether trained humans can apply the instrument, not whether a model can. It extended the instrument's coverage in ways the LLM-assisted phases had not, including codes reflecting culturally responsive teaching, world language instruction, and administrative finance, several of which correspond to user populations or uses that grew over the year the validation sample spans. It also disciplined the codebook's boundaries, since the calibration discussions that resolved disagreements doubled as tests of whether inclusion and exclusion criteria were writable and teachable. A pipeline that had stopped at the end of the selective phase would have delivered a plausible instrument with none of these warrants. We therefore see human application not as a check appended to an LLM method but as a constitutive phase of it.

## Limitations

Several limitations bound the account offered here. First, all data originate from one AI platform, and the codebook's domains reflect the professional activity that platform supports. Teams studying other populations or platforms should expect the instrument's structure to transfer more readily than its items. Second, the pipeline used specific commercial models available during the study period, and although we have argued that the human-in-the-loop structure buffers the design against model change, the particular candidate labels the models proposed are not reproducible in the strict sense. The prompts, the procedures, and the human decision points are what this article can make generalizable. Third, agreement among human coders, while adequate for multi-label coding at this codebook size and interpretable through its granularity gradient, leaves room for improvement, and the calibration protocol we describe is part of the method precisely because independent application without it would have been weaker. Finally, we observe educator requests and platform responses, not classroom implementation, and the codebook characterizes what educators sought from the AI rather than what they subsequently did with it.

## Conclusion

This article documented a human-LLM collaborative pipeline that adapted open, axial, and selective coding to develop, and then validated through systematic human coding, a hierarchical codebook capturing K-12 educator AI use across 45,000 messages. The pipeline's answer to the question of LLM assistance in qualitative analysis is conditional rather than binary. Language models contributed scale, consistency, and candidate structure in the phases where those qualities matter, while conceptual authority, boundary judgment, and the detection of coverage gaps remained the work of researchers with domain expertise, and the validation phase demonstrated that this human work was not residual but generative, extending the instrument in ways scaled annotation had not. We offer the procedures, prompts, and decision points reported here as a template that other qualitative research teams can adapt, contest, and improve as the methodological conversation about language models in qualitative inquiry develops.

### Data Availability

The educator-AI conversation data analyzed in this study contain potentially identifying professional context and cannot be shared per privacy agreement. The full codebook, the complete LLM prompts, and replication scripts with synthetic data templates are publicly available in the study's replication repository: https://osf.io/ju59q/overview?view_only=96afb71945044b8794e113f9763cc74a (anonymous repository for peer review). The coding results are available upon request.

**Appendix A. Prompt for Open Coding**

You are an expert educator. Your task is to identify K-12 educational components in the chunk of educator-AI dialogue.

You are provided with a request, its response, and corresponding follow-up request if available.

Request: {request}

Response: {response}

Follow-up Response: {follow-up response}

1. Content Subject Domain Area:

   Output the specific subject and its domain areas if applicable

   (e.g., "Math/Geometry", "Science/Physics", "Science/Motion").

2. Pedagogical strategies:

   Output specific strategies if applicable

   (e.g., "Formative Assessment", "Group Work", "Student Discourse", "Connect to Real-world").

3. Instructional Strategy (1-5 scale):

   - 5: The command clearly addresses instructional strategies.
   - 4: Touches on strategies but lacks depth.
   - 3: Implies strategies without a clear focus.
   - 2: Vaguely references instructional strategies.
   - 1: No instructional strategy mentioned.

4. Request for AI Tool Support:

   Output the specific support the teacher is asking for.

Format your responses for each item in short phrases.

Provide the results in the strict JSON structure:

```
{
  "Request": {request},
  "results": {
    "Content subject domain area": "subject area" or null,
    "Pedagogical strategies": "strategy" or null,
    "Instructional Strategy": score,
    "Request for AI Tool Support": "support" or null
  }
}
```

**Appendix B. Prompt for Open Coding (Expanded Schema)**

You are an expert educator. Your task is to identify K-12 educational components in the chunk of educator-AI dialogue.

You are provided with a request, its response, and corresponding follow-up request if available.

Request: {request}

Response: {response}

Followup Request: {followup_request}

1. Content Subject Domain Area:

Output the specific subject and its domain areas if applicable
(e.g., "Math/Geometry", "Science/Physics", "Science/Motion",
"Social Science").

2. Contextual Information:
Output components that are specifically contextualized for the teacher's
K-12 educational setting. For example, grade level, student demographics,
instructional challenges, student needs such as Special Education (IEP/504)
or English Language Learners (ELLs), or teacher needs.
If grade level is identified, format grades as:
"Grade_K", "Grade_1", "Grade_10", or "Grade_Hs".

3. Instructional Tasks:
Output specific instructional tasks or pedagogical strategies requested
(e.g., "Formative Assessment", "Group Work", "Student Discourse",
"Incorporate Real-world Example", "Reflective Learning",
"Deep Questions").

4. Instructional Strategy (1-3 scale):
- 3: The request clearly addresses instructional strategies.
- 2: Implies or touches on strategies but lacks a clear focus.
- 1: No instructional strategy mentioned or implied.

5. Learning Progression (1-3 scale):

Output keywords describing students' prior knowledge or ideas assumed or referenced in the request.

6. Pedagogical Frameworks:

Output any explicitly referenced pedagogical frameworks (e.g., "UDL", "World Cafe").

7. Request for AI Tool Support:

Output the specific AI support requested (e.g., translation, rubric creation).

8. Category:

Categorize the AI response using ONE label:

- Information
- Explanation
- Guidance
- Question
- Summarization

Notes:

- "Question" applies only to clarifying or contextual questions.
- Do NOT include generic follow-up questions.

Follow the output format requirement strictly.

DO NOT add additional text or reasoning.

Provide results in the strict JSON structure:

```
{
  "Results": {
    "Content subject domain area": "subject_area" or null,
    "Contextual Information": "educational_context" or null,
    "Instructional Tasks": "task" or null,
    "Instructional Strategy": score,
    "Learning Progression": "prior ideas" or null,
    "Pedagogical Frameworks": "framework" or null,
    "Request for AI Tool Support": "support" or null,
    "Response Category": "category label" or null
  }
}
```

**Appendix C. Prompt for Selective Coding**

The selective coding prompt embedded the preliminary codebook as a closed vocabulary. The model was instructed to match each trio against predefined label lists for subject area, grade level, educational context, content focus, and instructional practices, selecting “Other” with a brief specification when no predefined label applied, and to return results in a strict JSON structure. The complete prompt, including the full label lists, is available with the codebook as described in the Data Availability statement.